\documentclass[fleqn,10pt]{wlscirep}
\usepackage[utf8]{inputenc}
\usepackage[T1]{fontenc}
\usepackage{subcaption}
\usepackage[utf8]{inputenc}
\usepackage[T1]{fontenc}
\usepackage{mciteplus}

\usepackage{xcolor}

\definecolor{olivegreen}{rgb}{0.33333,.41961,0.18431}
\definecolor{forestgreen}{rgb}{0.13333,.5451,0.13333}
\definecolor{lightgrey}{rgb}{0.7,0.7,0.7}
\definecolor{verylightgrey}{rgb}{0.90,0.90,0.90}
\definecolor{veryverylightgrey}{rgb}{0.95,0.95,0.95}
\definecolor{grey}{rgb}{0.5,0.5,0.5}

\definecolor{headerblue}{HTML}{33367E}
\definecolor{unitednationsblue}{HTML}{4D88FF}

\definecolor{charcoal}{HTML}{36454F}
\definecolor{cinerous}{HTML}{98817B}
\definecolor{feldgrau}{HTML}{4D5D53}
\definecolor{glaucous}{HTML}{6082B6}
\definecolor{arsenic}{HTML}{3B444B}
\definecolor{xanadu}{HTML}{738678}

\definecolor{firebrick}{HTML}{B22222}
\definecolor{orangered}{HTML}{FF4500}
\definecolor{tomato}{HTML}{FF6347}

\definecolor{purpletaupe}{HTML}{3B444B}

\usepackage{textgreek}

\definecolor{todoblue}{RGB}{0, 91, 187}

\usepackage{hyperref}
\hypersetup{
    colorlinks=true,
    linkcolor=grey,
    filecolor=grey,
    urlcolor=grey,
    citecolor=todoblue,
    breaklinks=true,
}

\title{Limits of Individual Consent and Models of Distributed Consent in Online Social Networks}

\author[1]{Juniper Lovato}
\author[1,2,3]{Antoine Allard}
\author[1,4]{Randall Harp}
\author [1,5] {Jeremiah Onaolapo}
\author[1,2,5,*]{Laurent H\'ebert-Dufresne}
\affil[1]{Vermont Complex Systems Center, University of Vermont, Burlington, 05405, USA}
\affil[2]{D\'epartement de physique, de g\'enie physique et d’optique, Universit\'e Laval, Qu\'ebec, G1V~0A6, Canada}
\affil[3]{Centre interdisciplinaire en modélisation mathématique, Universit\'e Laval, Qu\'ebec (Qu\'ebec), G1V~0A6, Canada}
\affil[4]{Department of Philosophy, University of Vermont, Burlington, 05405, USA}
\affil[5]{Department of Computer Science, University of Vermont, Burlington, 05405, USA}

\affil[*]{laurent.hebert-dufresne@uvm.edu}


\begin{abstract}
Personal data are not discrete in socially-networked digital environments. A user who consents to allow access to their profile can expose the personal data of their network connections to non-consented access. Therefore, the traditional consent model (informed and individual) is not appropriate in social networks where informed consent may not be possible for all users affected by data processing and where information is distributed across users. Here, we outline the adequacy of consent for data transactions. Informed by the shortcomings of individual consent, we introduce both a platform-specific model of ``distributed consent'' and a cross-platform model of a ``consent passport.'' In both models, individuals and groups can coordinate by giving consent conditional on that of their network connections. We simulate the impact of these distributed consent models on the observability of social networks and find that low adoption would allow macroscopic subsets of networks to preserve their connectivity and privacy.
\end{abstract}
\begin{document}

\flushbottom
\maketitle
%
%
\thispagestyle{empty}


\section{Introduction}

One key focus of the burgeoning field of data ethics concerns how big data and networked systems challenge classic notions of privacy, bias, transparency, and consent \cite{leonard2018emerging}. In particular, the traditional privacy model, which relies on individual self-determination and individual consent, we argue, is no longer appropriate for the digital age. First, the traditional privacy model requires that consent be \emph{informed}, which may not be possible in the context of large data sets and complicated technologies. Second, the traditional privacy model presumes \emph{individual} control over personal information, even though the flow of information in networked systems precludes anyone from having such control over any piece of data. While the modern information environment shows both conditions as problematic, and while we briefly discuss the information condition, we focus most of our attention on the individuality condition.

Individual consent (by which we mean requiring that individual end-users consent in order for some action or outcome to be permissible) has many limitations---notably, we live in a highly networked \cite{marwick2014networked} and advanced technological society, where digital decisions and actions are interconnected and affect not just ourselves but our digital community as a whole. In a digital age, individual consent is flawed \cite{borgesius2015informed} and ineffectual when protected class data and social profiles can be easily inferred via our social networks \cite{garcia2017leaking, bagrow2019information, borgesius2016singling, matzner2014privacy, sarigol2014online}. The individual consent model works most effectively in a physical space with accepted boundary norms \cite{tufekci2008can, petronio1991communication}, linear contacts between two discrete parties, and no externalities. This, however, does not translate well to a digital realm where personal data boundaries are fuzzy and interwoven. The current over-reliance on individual consent online has also led to a negative externality of less legitimate consent due to consent desensitization, in part because users are now faced with a deluge of consent requests \cite{schermer2014crisis}. Thus, a new approach for data privacy and consent in this context is needed.

A new data privacy model will need to consider several factors: the networked virtual space that we occupy; integration of group consent; and a mechanism for distributed moral responsibility when data privacy is breached, or data are processed, combined, or manipulated in unethical manners \cite{floridi2016faultless}. In this paper, we will focus on distributed consent in particular and evaluate, in a mathematical model, its potential to increase online social networks' general privacy. We aim to cover the latter data privacy concerns in future work. In addition, future work could explore the potential for early adopters of distributed consent to influence their network neighbours (cascade effects towards a contagious taste for privacy).


\section{A Critique on the Adequacy of Individual Consent for Data Transactions}

We will call the means by which information is shared \emph{personal data transactions}.  Broadly speaking, a personal data transaction is any transaction in which one party gives or reveals personal information to another, so the category is vast; it includes every behaviour and every speech act that imparts information of some kind to another.

For our purposes here, we can narrow the broad category of personal data transactions in three ways. First, we are interested in those transactions for which the \emph{primary purpose of the actions or transaction is the transfer of information}. We will not attempt to give a complete conceptual framework here, but we can provide some indication of what we mean.  If $A$ gives $B$ money, and $B$ gives $A$ a shirt, information has been exchanged, but the primary purpose of $A$'s giving money and $B$'s giving the shirt was not the information exchange---it was the exchange of goods.  If $A$ asks $B$ how their day was and $B$ replies “pretty good,” information has been exchanged, but it is possible that the primary purpose was not the information exchange but rather the demonstration of caring, or the strengthening of solidarity.  So we will focus on data transactions where the primary purpose behind the transaction is the transfer of information itself---though it is important to know that the information might be valued because of further downstream uses of the information.  If $A$ asks $B$ about where $B$ grew up, and what $B$’s mother’s maiden name is, and the name of B’s childhood pet, then $B$’s responses would be data transactions---even if $A$’s ultimate reason in asking the questions were to get access to $B$’s online banking accounts.

Second, we are interested in those transactions which are about \emph{personal information}. As with the previous case, we cannot precisely conceptualize personal information here. The general idea is that personal information is information about an identifiable living person \cite{nissenbaum2009privacy}. If $A$ downloads all of the 1911 Eleventh Edition of the \textit{Encyclopaedia Britannica} and gives it to $B$, that is a data transaction in the broad sense but is likely not a transfer of personal information in the sense that we are interested in here (assuming that B is not an entry in the encyclopedia). If $A$ gives $B$ information about $A$’s whereabouts over the past 24 hours, or information about $A$’s food preferences, etc., that is personal information.  It is important to note that personal information need not be information about the person who is giving it to someone else; $A$ can give $B$ information about $C$'s food preferences or whereabouts, which would constitute a data transaction. Here again, the boundaries are difficult to lay out precisely.   

Where do the limits of personal data lie? Suppose $A$ were to give $B$ a copy of a biography of Barack Obama:  A’s giving $B$ the biography is outside of the scope of data transactions that we’re interested in because the personal information about Obama contained in the biography is presumed to be \emph{public}. Data transactions are valuable insofar as some party is gaining something of value, and discrete information that is previously known is not valuable. And that suggests a third way to restrict the domain of data transactions that we are interested in: we are concerned only with data transactions that deal with \emph{non-public information}. 

So when we talk about personal data transactions, we are talking about exchanges of personal information between two parties where the exchange of information is the (or a) primary purpose of the exchange, where the information is personal information, and where the information is non-public. Personal data transactions of this form make up a significant and increasing portion of our modern lives. The following examples are just two of many but highlight personal data exchanges that have the potential to be extremely impactful on our moral lives.

As one example, consider our use of social media.  When we make an account on social media, we are potentially engaging in three different kinds of personal data transactions: first, there are the data transactions between the person with the account (the “end-user”) and the social media company or platform.  Second, there are the data transactions between one end-user and other end-users on the platform between whom there are some network connections.  And third, there are the data transactions between end-users and third-party auxiliaries (individuals, apps, bots, etc.) that receive or process data in order to enhance the end-user’s experience of the social network.  As examples of these, consider Facebook: anyone creating a Facebook account shares with the site all of the information they intentionally put on the site (their profile information, their network connections/friends, their posts, etc.). There is also a host of information that they might not be aware they are putting on the site (their location, the amount of time they spend reading posts or watching ads, etc.).  The Facebook user also builds networks of social connections (nodes) on the platform. 

Selected information is shared among various nodes on the friendship network in accordance with the end-user’s preferences. End-user $A$ might share a lot of information with friends $B$ and $C$. User $A$ is close friends with $B$ and $C$ and may want to share information such as profile data, location data, the content of all of their posts, etc. User $A$ is an acquaintance with users $D$ and $E$ and may want data sharing with them to be limited to only some posts or only some photos, or no location information or information about $A$'s friend network. The third kind of information flow is that which is shared with third parties. For example, third parties might be  designers of games or quizzes that can be run on the Facebook platform, such that those third parties can then collect information Facebook possesses about the end-user playing the game. (Cambridge Analytica is an example of a third party who acquired information from end-users through games or apps on the Facebook platform; Facebook has since changed some of its policies on third-party data acquisition.)  Third-party websites can also allow users to sign in with their Facebook accounts, and those third parties can then get some user information when people use their Facebook accounts to log in, etc. While not every social media site offers the same options for how information is shared across those three modalities, having an account on any social media site entails sharing information in each of those three different ways.

\subsection{A Theory of Consent}
The fact that people have putatively consented to all of the personal data transactions that we identified above is supposed to be doing a lot of normative work: it takes information gathering actions that would have been impermissible and supposedly makes them permissible.
As we have seen, our modern lives are filled with personal data transactions.  And while it has been argued that the ubiquity of these personal data transactions is such as to entail that we are living within a  \emph{de facto} surveillance state \cite{zuboff:2019}, there is at least one important  \emph{prima facie} difference: surveillance states are typically imposed on subjects without their consent, so that personal information in a surveillance state is gathered without any consideration to the preferences the surveilled, whereas (it is claimed) we freely consent to the personal data transactions that we are subject to.  (Note that we are not here endorsing this argument; we present it merely in order to motivate our analysis of the role of consent in personal data transactions.)
In this way, consent for personal data transactions functions analogously to how consent functions in medical ethics and how consent functions in sexual ethics.

To borrow a phrase used by both Heidi Hurd and Larry Alexander, consent is a kind of ``moral magic'': ``it transforms acts from impermissible to permissible \cite{hurd:1996, alexander:1996}.''
This is true in generic ways; $A$'s entering $B$'s house can be either a trespass or a permissible visit based on whether $B$ has consented to $A$'s entering.  But it is particularly true in matters of sexual ethics, where the moral status of a sexual act crucially depends on whether the act is consented to at the time that it is performed \cite{dougherty2014fickle}. And it is equally true for medical ethics, where invasive medical procedures and treatments are impermissible unless consented to by the patient.

Regardless of the legality, it is reasonable to think that if $A$ were to persistently and systematically surveil $B$ (going through $B$'s trash, compiling every public record about $B$, recording all of $B$'s public movements), and $B$ were not a public figure who might be an appropriate target of community scrutiny, then that would be an impermissible form of information gathering. Of course, if $B$ were to consent to their information being gathered in this way by $A$ (say they are the willing subject of a documentary), then the information-gathering would thereby become permissible, and so consent would have the same kind of moral magic as it does in other contexts \cite{nissenbaum2004privacy}.

Likewise, personal information transactions open up the risk of one's autonomy or integrity being violated. After all, personal information is personal---and as such, it potentially enables others to identify them and predict or control one's behaviour in a way that compromises one's autonomy and capacity to act on one's intentions and one's own conception of the good. It is true that we share personal information with others around us all the time; the mere sharing of personal information does not compromise one's autonomy.  But we typically share personal information with those we trust to use the information correctly. We share information that is anodyne enough to not threaten our ability to pursue our own goals; we do not typically share personal information with those we know intend to use that information to circumvent our autonomy or act contrary to our interests. This is why privacy is important even for those who do not think themselves to have a strong taste for privacy: autonomy matters because it is the means by which we pursue the good, and privacy is connected to autonomy. Getting legitimate consent to personal data transactions helps to ensure that one's autonomy is safeguarded.

Finally, personal data transactions often involve commercial parties, either as one of the parties to the data transaction (as happens when you upload your information to Facebook), or as the platform in which data transaction occurs (as when you share your data with your friends through Facebook). Commercial parties have an interest in limiting their liability. To this end, obtaining putative consent to acquire and process data helps limit the legal liability they face. This is especially true given that there is very little case law governing personal data transactions \cite{solove2004digital}. Companies and commercial agents have a vested interest in preventing the end users of social media sites and data technologies from complaining, and so obtaining consent helps to support the argument that the end users have, indeed, forfeited their right to complain about any consequences that arise from the personal data transactions.

\subsection{Consent in Data Transactions}
Before talking about the limitations of the individual consent model for personal data transactions, it is worth addressing one question: namely, why would we ever have thought that consent was relevant for data transactions at all?  After all, the argument goes, data transactions are just one species of transaction.  And transactions are necessarily mutually consensual; if they were not, they would not be transactions.  An exchange in which $A$ gets a beer and $B$ gets a dollar is a transaction if they both agree, but it is robbery or coercion if either $A$ or $B$ do not consent to the exchange.  (The fact that both $A$ and $B$ receive something is irrelevant, of course; if $A$ breaks into $B$'s house and steals $B$'s property, it is no less a robbery just because $A$ left something behind in exchange.)  Likewise, the objection goes, if $A$ and $B$ are engaged in a personal data transaction, then it is irrelevant to ask whether the transaction is consensual; if it were not, it would not be a personal data transaction but would instead be a data theft or something similar.  If this is right, it is as unnecessary to ask whether a data transaction is consensual as it would be for the cashier at a clothing store to explicitly ask every patron whether they consent to trade their money for the clothes they wish to buy.

One reply to this objection is to say that we actually \emph{do} care about obtaining consent for some exchanges, precisely because we want to ensure that the exchange is a ``transaction'' rather than a ``theft.''  When the stakes are high, and there is a possibility for future risk, in other words, we do seek to clarify the consensual nature of the exchange, but when the stakes are low and there is a low perception of risk, we are less concerned \cite{skirpan2018what, acquisti_what_2013, farmer2009hyperbolic}.  And this reply is right, but we think that there is an important point that is in danger of being obscured.  In the case of ordinary transactions, there is less danger of the transaction being non-consensual precisely because there is little danger of the parties to the exchange not knowing what they are exchanging; when $A$ gives $B$ money and $B$ gives $A$ a shirt, both parties are aware that the transaction happened because both parties have clearly gained something and have clearly lost something.  Moreover, there are positive actions that $A$ and $B$ both perform, without which the transaction cannot take place; $A$ must hand over the money and $B$ must hand over the shirt.  But we can easily imagine transactions in which these conditions are not met---perhaps, for instance, it is not clear among the parties exactly what has been gained and what has been lost after a transaction.  (One might think of the ``sale" of the island of Manhattan by Native people that Peter Minuit orchestrated on behalf of the Dutch; what exactly is one selling if the land is still there after the transaction is done?)  Or imagine transactions for which no positive actions need to be taken, like arrangements that automatically deduct money from a bank account or other store of value without anyone needing to do anything.  It is certainly not difficult to sway our intuitions towards thinking of these ``transactions" as theft, or at a minimum theft-adjacent.

The same issues arise with data transactions (whether legitimate or not).  We might wonder whether the exchange of data between $A$ and $B$ is actually a transaction if one or both parties are not clear on exactly what has been gained or lost---but, as we will see, this is a common feature of most modern personal data exchanges.  It is certainly not like a transaction for a shirt, where one minute you have a dollar and the next minute you do not; after a personal data exchange is complete, you have just as much of your own personal data as you started with.  Personal data exchanges often do not require any positive action on behalf of the parties; we lose data in personal data exchanges all the time, through no action of our own.  So while it is fine to say that true transactions are consensual, it is also true that not every data exchange is a data transaction in that strict sense. Given that, it is reasonable to ask for explicit consent to data exchanges so that all parties are confident that it is a data transaction and not a mere exchange.

It is a reasonable strategy, but it is nevertheless a failed one.  As we will see, there are systematic reasons why personal data exchanges cannot be justified by individual consent.

\subsection{A General Overview of the Problems with Individual Consent}
We can now provide a very general overview of the problems with individual informed consent when applied to data transactions. Note that we are assuming for the sake of this discussion that the relevant data are, in fact, properly subject to control by individuals.  This is a problematic assumption; data very often implicate multiple individuals or are otherwise `co-owned' and thus are not properly the sorts of things that individual consent can govern.  This is an important point which requires a fuller discussion, but we make this simplifying assumption here because social networks cannot function in anything like their current form without it.

Consent, in this context, should not be mistaken for a state of mind or an attitudinal event \cite{kleinig1982ethics,westen2017logic} and it must meet certain criteria in order to be considered a valid. The legitimacy of consent hinges on a number of criteria \cite{faden1986history}:
\begin{enumerate}
    \item the subject has sufficient accurate information and understands the nature of the agreement,
    \item the agreement is entered into without coercion,
    \item the agreement is entered into knowingly and intentionally,
    \item the agreement authorizes a specific course of action.
\end{enumerate}

In the context of personal data transactions, digital consent also rests on the four aforementioned criteria as well as the user agreeing to the specified service terms and privacy policy outlined by the data processor. Notably, the four criteria listed above fail in the context of personal data transactions and classic Privacy Policies and Terms of Service (ToS) agreements.

First, most users entering into consent agreements know very little about data processing or the risks associated with handing over their data. The dense legal and technical nature of ToS agreements task non-experts to consent to something they do not understand \cite{custers2013informed}. This dynamic takes advantage of asymmetry in technical and legal knowledge. 

Second, it is difficult to opt-out of these services since online platforms are an important social ecology where people form personhood, maintain personal relationships, and build valuable networked counter-publics \cite{fraser1990rethinking,jackson2015hijacking,jackson2016digital, rouvroy2009right}. Yet, there is little to no power on the part of the individual to negotiate the ToS with these companies, as consent in these ToS are typically presented on a take-it-or-leave-it basis and offer no conditions of choice \cite{schwartz2000internet, obar2017clickwrap}. Online privacy then turns into an unfortunate social optimization problem \cite{tufekci2008can}, where the user must choose between the pressures of disclosing too much personal information (being digitally crowded) and being socially isolated \cite{altman1977privacy}. 

Third, the volume of consent requests a user faces has led to a troublesome externality where the user is fatigued and habitually agrees to everything due to consent desensitization \cite{custers2013informed}. This delegitimizes the premise that each act of putative consent actually reflects the individual user's autonomous judgment.

Fourth, the language in ToS agreements are typically so broad and open-ended that data processors have the flexibility to manipulate the data in many ways. The consent scope cannot be so broad as to allow actions that the user could not have considered or would otherwise not have consented to. A properly limited scope of consent also implies that there should be some mechanism for a user to check if their data are indeed following the agreed-upon course of action. However, data processors often make it very difficult \cite{lapowsky19}, if not impossible, to track personal data, know what they have collected or how it is being processed, and hold them accountable for misuse \cite{solove2004digital, cate1999principles}. 

Perhaps more importantly, a significant concern with the individual consent model is that personal data, in this context, are distributed information that contains information about more than a single individual and spans a wider communication boundary than the user is be aware \cite{petronio1991communication}. A fundamental assumption for individual consent is that the user has power over their personal data and can trade their personal privacy in exchange for using an online service \cite{cohen2000examined}. In reality, these data may not be wholly the individual's, and therefore, it is not appropriate for the individual to act alone in controlling the course of action or the flow of these data. Perhaps the first step to understanding the impact of this issue in an online social media context is to understand how different levels of consent impact the flow of networked data and observability in the first place. It will be important for us to understand if the network effect indeed has a strong influence on privacy in order to justify group level consent settings.

\section{A Modelling Framework for distributed Consent}

\subsection{A Threat Model for Leaky Individual Data in Social Networks}
\label{sec:threatmodel}

The densely interconnected nature of online social ecology creates a significant problem with the model of individual consent. When users share personal information online, they are also leaking personal information about others in their social network (digital or otherwise) \cite{garcia2017leaking,bagrow2019information}. According to Bagrow et al. ``due to the social flow of information, we estimate that approximately 95\% of the potential predictive accuracy attainable for an individual is available within the social ties of that individual only, without requiring the individual’s data \cite{bagrow2019information}.''


One example of leaky data is when a user attempts to signs onto a new online service, they may be prompted to skip the hassle of entering their personal information manually and instead opt to use an existing account to act as a secured access delegation \cite{wang2011third} in order to gain quicker access to the new third-party online service. In turn, the online service can ask to gain access to the user's online social network, phone or email contacts, location, and other personal data. Through these leaky data, third parties can be granted access to a wealth of knowledge about people who never consented to share their information with that particular service.

Similarly, attackers can breach social accounts via various methods including phishing attacks, malware, and data breaches \cite{dhamija2006phishing,stone2009botnet,halfond2006sqlinject}. They can also create fake social accounts and then use those fake accounts to befriend real accounts. To extend their reach, those attackers can then monitor the activity of other accounts that are directly connected to the compromised or fake accounts. By leveraging network effects, attackers can further indirectly observe the social activity of groups of accounts that are several hops away from the captive accounts, starting with accounts that are one hop away.

Leveraging those captive accounts, the attackers traverse the social graph, or segments of it, and record profile information and social activity that would later be used in future phishing and spam attacks, influence manipulation attempts \cite{wylie2019mindf}, and disinformation campaigns \cite{asmolov2018disinfo}, among others. Given the massive size and connected nature of online social networks, the potential reach of attackers and the resulting harm both have the capacity to rise to catastrophic levels. Section \ref{sec:implications} discusses examples of real-world incidents that demonstrate the severity of this problem.



\begin{figure*}
    \centering
    \begin{subfigure}{0.3\linewidth}
    \includegraphics[trim={0 0 0 0.2cm},clip,width=\linewidth]{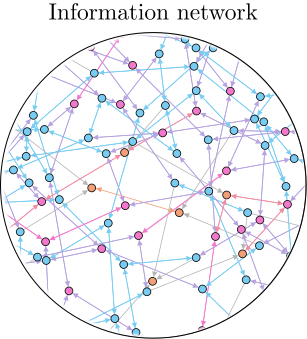}
    \caption{Information network}\label{fig:1a}
    \end{subfigure}
    \hspace{0.7cm}
    \begin{subfigure}{0.3\linewidth}
    \includegraphics[trim={0 0 0 0.2cm},clip,width=\linewidth]{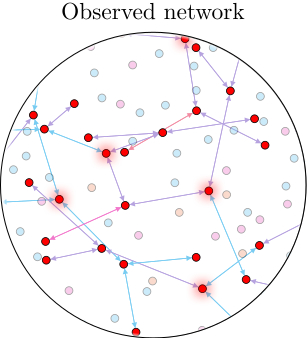}
    \caption{Observed network}\label{fig:1b}
    \end{subfigure}
    \hspace{0.7cm}
    \begin{subfigure}{0.3\linewidth}
    \includegraphics[trim={0 0 0 0.2cm},clip,width=\linewidth]{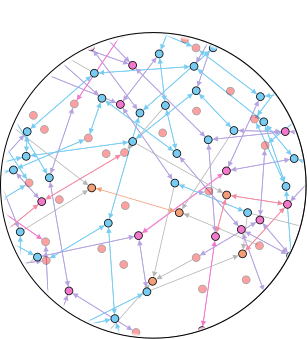}
    \caption{Protected network}\label{fig:1c}
    \end{subfigure}
    \caption{(a) Cartoon of information flow across a network with our basic implementation of distributed consent. Blue nodes have the lowest security settings, and are susceptible to surveillance from third-party applications or websites. Purple nodes have stricter security settings but share their posts and, therefore, data with all their neighbours. Orange nodes follow a distributed consent model and only share their data with purple nodes or other orange nodes. (b) The same network where a handful of low-security accounts are directly observed by a third party, highlighted in red with shading. All nodes sharing their data with directly observed accounts are de facto observed as well, and are also shown in red. Nodes at a distance $L>1$ can also be observed if the third party leverages some statistical procedure, in this case inferring data up to a distance of two from directly observed nodes. (c) We show the remaining unobserved, or protected, component network. Note that, under this observability process, orange nodes who follow a distributed consent model are much less likely to be observed than nodes following traditional individual consent options. Importantly, they also help protect some of their neighbours with standard security settings.}
     \hspace{1.5cm}
  \label{fig:teaser}
\end{figure*}

\subsection{Results: A Model of Distributed Consent and Network Observability}

To account for the distributed nature of personal data (i.e. the distributed online self), we consider a simple model of distributed consent. Imagine a social network platform where individuals have the following privacy options:
\begin{itemize}
    \item[0.] Individuals share their data with all their connections and are vulnerable to third-party surveillance (similar to Facebook accounts with access for ``Apps, Websites and Games'' turned on).
    \item[1.] Individuals share their data with all their connections but are not directly vulnerable to third-party surveillance.
    \item[2.] Individuals only share their data with their connections whose privacy level are set at least to 1.
    \item[N.] Individuals only share their data with their connections whose privacy level are set at least to $N-1$.
\end{itemize}

Options 2 and greater are currently unavailable on popular social media platforms (though they could be an attractive setting to adopt if the platforms wish to address privacy concerns and keep users) but are a first-order implementation of what we call distributed consent. Individuals who pick this option are stating that they want to be part of a local group that agrees on minimal privacy settings. It is a consent that is conditional on the consent of their neighbours in the network structure. 

Imagine now that a third party wishes to observe this population, either by releasing a surveillance application on the social network or by explicitly gaining control of their accounts through similar malware. Say they can directly observe a fraction $\varphi$ of individuals with privacy level set to 0 through this attack. They can then leverage these accounts to access neighbouring individuals' data with the privacy level set to 1 or 0, therefore using the network structure to indirectly observe more nodes. They can further leverage all of these data to infer information about other individuals further away in the network, for example, through statistical methods, facial recognition, other data sets, etc.

Deep surveillance processes are relevant to other types of network systems where it is possible to indirectly observe nodes within a certain distance of directly observed nodes. Deep surveillance allows us, for example, to monitor an entire power grid without monitoring the voltage and line currents everywhere in the system \cite{yang2012network}. It was recently shown that the surveillance process itself can be generally modelled through the concept of depth-L percolation \cite{allard2014coexistence}: Monitoring an individual allows one to monitor their neighbours up to L hops away. Depth-0 percolation is a well-studied process known as site percolation. The third-party would then be observing the network without the help of any inference method and by ignoring its network structure. With depth-1 percolation, they would observe nodes either directly or indirectly by observing neighbours of directly observed nodes, e.g., by simply observing their data feed or timeline. Depth-2 percolation would allow one to observe not only directly monitored nodes, but also their neighbours and neighbours’ neighbours, e.g., through statistical inference \cite{bagrow2019information}---and so on, with deeper observation requiring increasingly advanced methods. The model is illustrated in Fig.~\ref{fig:teaser} and detailed in our Methods section.

To study the interplay of consent and observability on social networks, we combine our models of distributed consent and depth-L percolation on subsets of Facebook friendship data informed by empirical work on demographic population's taste for privacy. We assume that one-third of the population has a taste for privacy \cite{lewis2008taste}. These data are anonymized with all metadata removed, and are simply used to capture the density and heterogeneity of real online network platforms. We use distributed consent with security level up to $N=2$ and an observation process with $L=2$ (observing nodes two hops away from the compromised account). As we will see, those values mean that the third-party is more sophisticated than our distributed consent mechanism and no one is \textit{guaranteed} to be unobservable. We then set 1\% of accounts with the lowest security setting to be compromised and directly observed such that that between 90\% and 100\% of the population will be observed given the default security settings. We then ask to what extent distributed consent can preserve individual privacy even when a large fraction of nodes can be directly observed and third-parties can infer data of unobserved neighbours. How widely should distributed consent be adopted to guarantee connectivity \textit{and} privacy of secure accounts? 

\begin{figure*}[t!]
\centering
    \includegraphics[width=\linewidth]{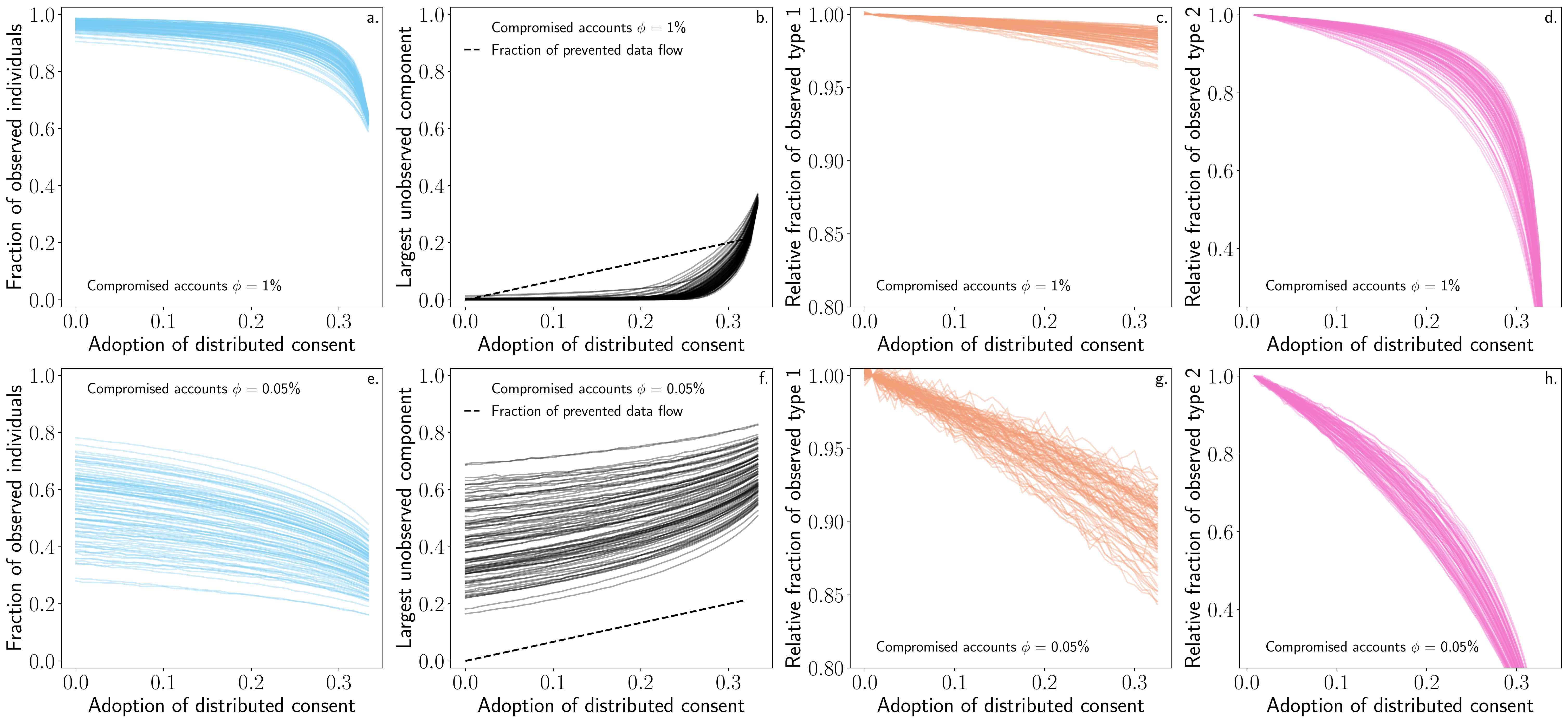}
    \caption{We use the anonymized Facebook100 dataset~\cite{traud2012social}. We assume that one-third of the population has a taste for privacy \cite{lewis2008taste}, split between security options 1 and 2 (i.e., classic or distributed consent) according to the adoption rate of distributed consent shown on the horizontal axis, while the remaining two thirds will use the default setting with the lowest security, option 0. We set 1\% (top row) or 0.05\% (bottom row) of accounts with security option 0 to be directly observable by a third-party app, which can also observe neighbours up to two hops away in the network. These values are chosen to model attacks that observe nearly the entire population (top row) or about a half on it (bottom row). We then vary the adoption rate and measure the total fraction of observed accounts (blue curves), the relative size of the largest unobserved connected component (black curves), and the fraction of observed individuals with security option 1 (orange curves) or two (pink curves).}
    \label{fig:emergence_of_components}
\end{figure*}

\begin{figure*}[t!]
\centering
    \includegraphics[width=\linewidth]{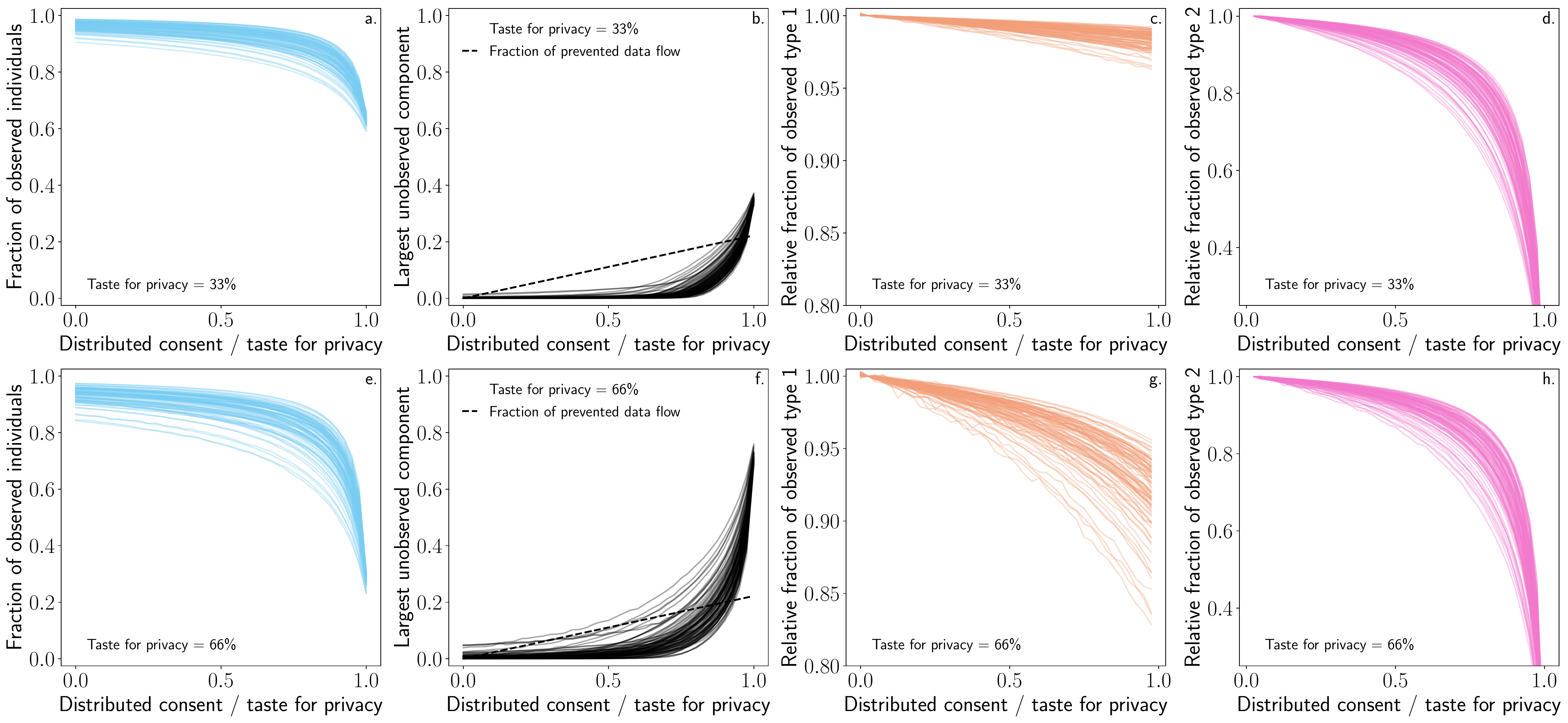}
    \caption{We reproduce results from Fig. \ref{fig:emergence_of_components} but using populations with different taste for privacy. The results in the top row of this figure match the top row of Fig. \ref{fig:emergence_of_components} but are now shown as a function of the nodes with a taste for privacy (security level greater than 0) that opt for distributed consent (security level greater than 1). We then change the fraction of population with a taste for privacy from one 33\% (top row) to 66\% (bottom row). Qualitatively, the results are very similar. The key quantity that drives the macroscopic effects of distributed consent is therefore not its total adoption but its relative adoption within individuals that cannot be directly observed. }
    \label{fig:emergence_of_components2}
\end{figure*}

The results of our simulations are shown in Fig.~\ref{fig:emergence_of_components} and Fig.~\ref{fig:emergence_of_components2}. We focus on the number of observed nodes of different security levels, and on the size of the giant component of unobserved nodes. This last quantity refers to the size of the largest subpopulation of accounts who are not observed and maintain connected pathways of any lengths between one another, thus preserving both their individual privacy and the global connectivity objective of the social network. In classic percolation theory, only one giant component can span the entire system \cite{stauffer2018introduction}, meaning only one subset of nodes can scale with the total size of the social network. Yet, from recent results on network observability \cite{allard2014coexistence}, we also know that giant \textit{observed} components can co-exist with giant \textit{unobserved}  components. This is where distributed consent can play a large role: Even if a third-party surveillance system scales with the size of the social network, it is theoretically possible for accounts to maintain their individual privacy and global connectivity at scale.

Figure \ref{fig:emergence_of_components} shows results of our model in populations facing either a very strong attack (1\% of compromised accounts, chosen to observe almost the full population) or a more modest attack (0.05\% of compromised accounts, chosen to observe about 50\% observation).
Against a strong attack, we find that while extremely low adoption levels of distributed consent have little impact on the observability of the system, moderate adoption (roughly 1 in 5 users) can lead to a transition where observability now drops sharply with adoption of distributed consent; see Fig.~\ref{fig:emergence_of_components}(a\&e). At low adoption rate of distributed consent, there are few unobserved nodes, all are mostly disconnected from each other and therefore observable through compromised neighbours. At higher levels of adoption rate, the system transitions to an unobservable and connected phase where privacy can co-exist with connectedness and information flow; see Fig.~\ref{fig:emergence_of_components}(b\&f). With large scale adoption of distributed consent (say one-third of users), we find that close to half of all accounts are now protected even against very strong attacks, even if their privacy settings only prevent about 22\% of data flow around them. 

Against the more modest attack, increases in adoption of distributed consent lead to an increase in protection which is smaller but smoother and more reliable. With 33\% adoption, populations can halve the size of their observed population and conversely double the size of their unobserved component.

To understand these results, notice that any user with privacy settings set to a greater value than the percolation depth will be unobservable. If we had set simulated naive attackers who observe at a depth $L=1$ only, adopters of distributed consent would have been fully protected. Indeed, users using a security setting $N$ will only share their data with users using settings of $N-1$ or more, and this statement holds for all $N$. We thus know that users using setting $N$ will be at least $N$ steps away from users using the lowest setting, which are the only directly observable nodes. Users with security level set to $1 < N < L$ can however be observed indirectly through their relationships. At low levels of adoption of distributed consent, a large amount of luck is required to remain unobservable (e.g., having zero connections with low-security users). At higher levels of adoption, users of distributed consent connect to, and therefore protect, one another. These connections are, however, localized and do not spread throughout the entire system. We find that when roughly 25\% of nodes with a taste for privacy adopt distributed consent, a large macroscopic component of connected unobservable nodes emerge even against the strongest attack. This component reflects a parallel, protected community that is unobservable but still connected to the rest of the social networks.

Despite the fact that a phase transition in connected unobservable nodes occurs at fairly low level of distributed consent adoption and that these nodes provide secondary protection to other users, pervasive adoption of group consent is required to fully protect a network. Again, all it takes for one vulnerable node to be indirectly observed is a single observable neighbour. Because of this and because of the density of most online networks platforms, it is extremely hard to completely protect vulnerable nodes even if distributed consent provides some secondary protection to all nodes. We thus see the coexistence of both observed and unobserved connected components at medium adoption level of distributed consent. Interestingly, these components are interconnected, with data flowing both ways across observable and unobservable components, yet the users in the latter remain fully protected from statistical inference of their data.

Importantly, the macroscopic but unobservable component that we see emerge with increased adoption of distributed consent does not only contain adopters of distributed consent. Early adopters of distributed consent provide some low amount of \textit{herd privacy} to the population, protecting otherwise vulnerable users; see Fig.~\ref{fig:emergence_of_components}(c\&g). Users with lower privacy settings can thus also benefit since the adoption of distributed consent in one's neighbourhood reduces the probability that one of their neighbours is directly or indirectly observed, thereby reducing the probability that they are themselves observed. However, as long as a majority of users rely on the default lax security settings, this effect will be limited as a single compromised neighbour is sufficient to observe a node. We do find, however, much stronger herd privacy effects against more moderate attacks. And similar effects provide non-linear returns on relative protection of distributed consent users, Fig.~\ref{fig:emergence_of_components}(d\&h), consistent with our previous observation of the emergence of an unobserved component.

Finally, in Fig.~\ref{fig:emergence_of_components2}, we reproduce the large attack against a population with a stronger taste for privacy, and compare our previous results with those obtained by halving the fraction of unprotected nodes (two thirds to one third). Unsurprisingly, the stronger the taste for privacy, the stronger the effects of distributed consent; both at the macroscopic level (the fraction of observed nodes and the size of the unobserved component in panels a, b, e and f) and at the microscopic level (herd privacy effects shown in panels c, d, g, and h). To make the comparison easier, we plot all quantities against the fraction of users with a taste for privacy (any security level other than the lowest) that opt for distributed consent. This ratio collapses all results on similar curves and therefore appears to be the critical quantity in determining the privacy level of a population.

\begin{figure*}[t!]
\centering
    \includegraphics[width=\linewidth]{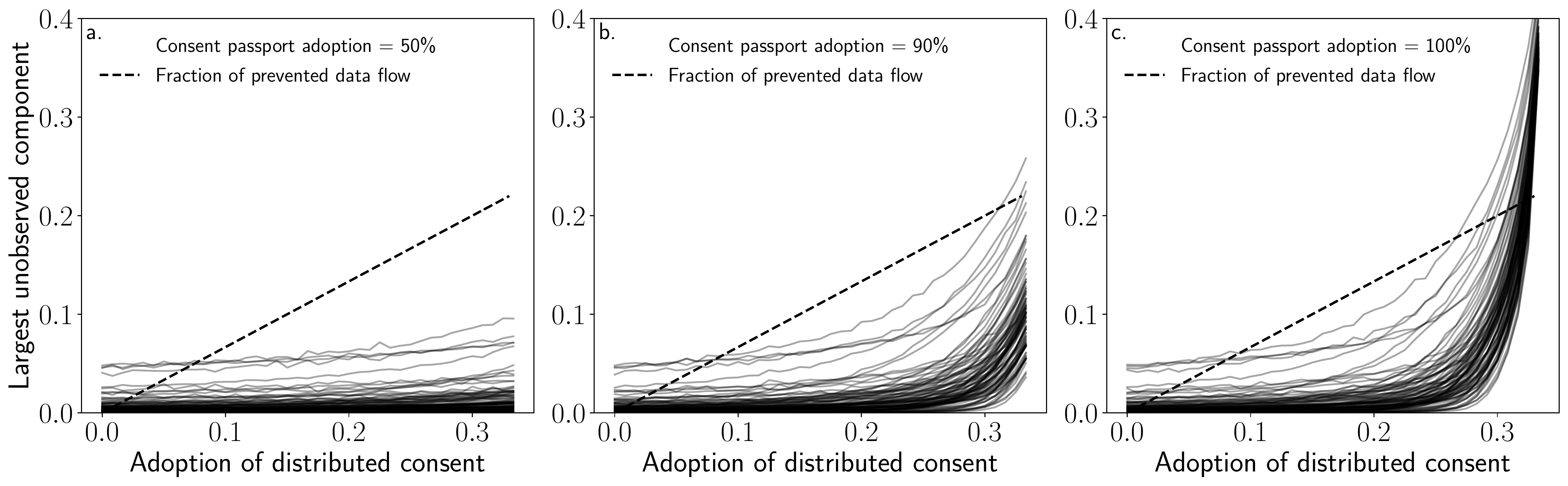}
    \caption{We again use the anonymized Facebook100 dataset~\cite{traud2012social}. We now create a multilayer network by doubling the original data, mimicking a two-platform ecosystem. We use the same parameters as in Fig.~\ref{fig:emergence_of_components}, assuming that for each platform only one third of accounts have changed their default security setting to options 1 or 2 (i.e., classic or distributed consent) according to the adoption rate of distributed consent shown on the horizontal axis. The remaining two thirds will use the default setting with lowest security, option 0. To account for the doubled network density and for the fact that users can be observed on either platform, we now set 0.25\% of accounts with security option 0 to be directly observable by a third-party app which can also observe neighbours up to two hops away through any layer of the network. We then vary the adoption rate of distributed consent (horizontal axis) and, within that subset of the population, vary the adoption of consent passport (different panels). In the resulting systems, we measure the relative size of the largest unobserved connected component. Randomly protecting users on a single platform does not protect anyone if spyware can jump network layers, but distributed consent coordinated across platforms through a consent passport can restore our ability to create an unobservable component within the multilayer ecosystem.}
    \label{fig:importance_of_passports}
\end{figure*}

\subsection{Results: A Model of Coordinated Consent Across Platforms}
 
Currently, there is a David and Goliath problem \cite{loos2016wanted}; the inequality of knowledge and power between the user and the data collector functionally takes away the individual's ability to realistically control their personal information. Part of the problem we believe is that the terms of agreement  are predicated solely on the platform's norms. Users do not have the option to opt-out and pay for the service in exchange for limiting information flow for most online social networks. There is a significant cost on the user's side to read and understand all ToS agreements presented to them \cite{mcdonald2008cost}. 

There is little to no power on the part of the individual to negotiate the ToS with tech companies. Online consent in its current state creates an unfortunate social optimization problem, where the user must choose between the pressures of disclosing too much personal information (being digitally crowded) and being socially isolated \cite{altman1977privacy, barnes2006privacy}. Moreover, this also implies that most platforms have little power to change the digital ecosystem on their own if users can be exposed through other platforms with laxer security and privacy policies. Multiple platforms create a multilayer network where information flows not only through social connections but across platforms through different layers of the network. This complex network structure exposes users through whichever platform offers them the least privacy.

Our second model focuses on coordinated consent across platforms. Inspired by previous work on automated ToS tools \cite{lippi2019claudette, pearson2014taking, santos2021consent, guarino2021machine, bannihatti2020finding, oltramari2018privonto, micklitz2017empire}, we envision a \textit{consent passport} model where instead of relying on every platform to offer advanced privacy settings, and therefore asking users to adjust their settings on every platform individually, users could use a consent passport stating their desired privacy settings before they join a platform. This could take the form of a key enabling a user to set their privacy baseline criteria based on their taste for privacy. The key will act to shift the burden \cite{colnago2020informing, grunewald2021tilt, sloan2014beyond} away from the user to read and understand the legalese of the platform's privacy policy and ToS. The consent key would restrict login and present a warning \cite{calo2011against} when the user attempts to enter sites that do not meet a minimum privacy criteria in their ToS and privacy settings. This key is intended for users who have a taste for privacy but do not want to fall prey to consent desensitization. The consent passport will need to be dynamic in nature so that users can remain autonomous in their decision-making and easily opt to enter a platform with discordant privacy settings if they have trust in the site or have decided that the privacy cost is worth the risk.  Importantly, this ensures that a given user's security settings will be coordinated across platforms, which might be the only way to protect them in a complex multilayer ecosystem confidently. 

To include a consent passport within our simulation model, we turn our network data into a multilayer infrastructure by duplicating the Facebook networks' structure to represent two platforms where users are randomly assigned a security setting, independently on each platform. However, a subset of users adopt a consent passport which guarantees that they will follow our previous model of distributed consent on \textit{both} platforms. We again use depth-L percolation with $L=2$ in the resulting systems to simulate a third-party's ability to observe network users. As a worst-case scenario, we allow the depth-L percolation process to be able to freely jump between layers of the networks; e.g., observing a Facebook neighbour of an individual directly observed through Instagram. As detailed in our Methods section, we classify any user as observed if they are observed on either platform.

Figure~\ref{fig:importance_of_passports} shows that distributed consent alone can not protect you if your security settings are not coordinated. Indeed, we parametrize the system such that the networks are roughly equally observable, and yet there is no emergence of a giant unobservable component even with medium adoption of consent passports (50\% adoption among distributed consent users, left panel). It is only at very high adoption of consent passports that we start seeing a non-linear benefit in unobservability (90\% adoption, middle panel) and only at near perfect adoption that we protect more users than we prevent data flow (95\% and above, right panel). These results demonstrate that multi-platform ecosystems are a much more complex beast to protect; users participating in multiple platforms are only as secure as their weakest security settings. Coordinating privacy settings across platforms are therefore a critical part of the solution. We discuss this further in Section \ref{sec:discussion}. 

While these results illustrate how mathematical and computational models can contribute to the study of consent and privacy policies, it is critical to keep all of their assumptions and approximations in mind before drawing conclusions about real-world applications. In the case of our model, we have assumed that all users generate and share similar amounts of data, that all users with a given security setting are as susceptible to privacy breaches, and that security settings are uncorrelated with the connectivity in the network. In practice, one might imagine that high profile accounts tend to have higher security settings but that they might also face more frequent attacks. Accounting for correlations between network structures, taste for privacy and susceptibility to breaches could be included in the model, but these mechanisms first require further empirical studies.

\section{Discussion}
\subsection{Implications}
\label{sec:implications}

In recent real-world incidents, attackers reportedly leveraged online social networks to target groups of people. In 2020, BBC News reported that a foreign intelligence agent allegedly used LinkedIn, a large online social network, to locate and befriend ``former US government and military employees'' \cite{BBC2020intel}. Additionally, the same BBC article reported that Germany's intelligence agency stated that foreign agents ``used LinkedIn to target at least 10,000 Germans'' in 2017. Yet another example comes to mind: the Cambridge Analytica case \cite{wylie2019mindf} in which political entities made attempts to sway the political stance of groups of people via Facebook, another large online social network.

In view of these real-world examples, the need for a model of distributed consent as presented in this paper becomes more apparent. In other words, the broad reach of attackers within the threat model presented in Section \ref{sec:threatmodel} could be restricted with the deployment of the model of distributed consent. Although the proposed model would not completely stop the attackers, it offers a better level of protection to users of online social accounts than the status quo. In view of this, it is our hope that online social network platforms would consider and adopt the distributed consent model. We also hope that policy makers would actively push for the adoption of the model to help make online social networks safer for all users.

\subsection{Conclusion}
\label{sec:discussion}
Altogether, in this work, we provided a philosophical critique of individual consent in the context of data transactions and used a modelling framework to suggest potential solutions.

As part of our philosophical critique, we listed four criteria for the legitimacy of informed consent. We argued that none of the four criteria are met by individual consent within online media's complex ecosystem. Further, a fundamental problem is that if personal data are distributed across individuals, so should be their consent.

Our results based on computational models and simulations suggest that even the simplest implementation of distributed consent could allow users to protect themselves and the flow of their data in the network. They do so by consenting to share their data conditionally on the consent or security settings of their contacts, thereby not sharing their data with users who might, in turn, make them available to third parties. This simple condition allows users to authorize a specific course of action for their own personal data (criterion 4). 

While this protection disconnects them from some other users, only a relatively low level of adoption of distributed consent is required to create a connected, macroscopic, sub-system within existing online network platforms. This sub-system consists of different individuals, including some that are granted secondary protection despite their low-security settings, and remains connected to the rest of the system such that information still flows throughout the entire population of users. Via this protected sub-system, distributed consent removes the \emph{de facto} coercion (criterion 2) involved in forcing individuals to choose between relinquishing control of their data or simply not participating in a platform.

Beyond the actual protection mechanism, this new consent model may also have interesting behavioural impacts on the users. Exposing users to this type of coordinated privacy setting might prompt them to reflect on their personal data's distributed nature and its flow through online media. This realization may encourage users to openly voice their social boundaries to their social network or restrict sending sensitive information to social neighbours who do not share their taste for privacy \cite{lewis2008taste}. Imagine a user publishing a post to their social network, before enacting the new privacy settings, urging those who want to remain connected to change their settings as well. Beyond the utility of limiting the social network's observability, this measure could also serve as an important educational tool on the interconnectedness of personal data (criterion 1). 

In a modest form, distributed consent could allow concerned users to protect themselves without fully leaving a platform. It would also let platforms maintain a large critical mass of observable users that chose to remain vulnerable and who are not granted sufficient protection through their contacts. 

That being said, legitimate consent criterion 1 (understanding the consent agreement) and criterion 3 (or consent fatigue) remain an issue that will need additional consideration in future work. An important caveat is that a useful implementations of distributed consent might require platforms provide additional education regarding data privacy. Regarding Criterion 1 and 3, the consent passport does attempt to shift the agreement's burden to platforms rather than users. In doing so, it may provide additional protection in complex multi-platform ecosystems. There are many types of privacy violations that are not solved by distributed consent.  These data are still leaky; individual users can still aggregate information about their neighbours without their direct consent. And finally, while the distributed consent model goes beyond the strict individuality of the traditional privacy model, it does so modestly; it still models the agents, choices, and values as fundamentally individual. Obviously, there is no silver bullet to solve this multi-scale complex problem; data privacy is a significant societal issue with multi-level interdependencies that need to be considered thoughtfully and ethically. Much work remains to be done in this area. 

Future work should extend to look at possible collective behaviour around the adoption of new models of consent. Indeed, the greatest hurdle to herd-like immunity against network observability is our assumption that only one-third of the population has a taste for privacy such that two-thirds of users will never deviate from the default lax security settings. Users signalling their adoption of distributed consent and potentially influencing their network neighbours to do the same could then spark a contagious taste for privacy whose co-evolution with observability could be modelled using tools from network epidemiology \cite{pastor2015epidemic}.

Beyond new notions of consent, effective data privacy measures will need to take a systems-level approach and integrate a mechanism for distributed moral responsibility \cite{floridi2016faultless} that will simultaneously involve both top-down and bottom-up interventions. Doing so will involve a synergy between increased governmental and professional regulation, technological intervention, distributed consent, and citizens' empowerment. Increasing data privacy and protection is not only an essential public service but a democratic imperative \cite{fraser1990rethinking,rouvroy2009right,dutt2018senator}. Access to data privacy and protection is a growing global issue \cite{mba2017flipping}, and it must be investigated through further multidisciplinary collaboration.




\section*{Methods}
{
\small
\textbf{Data.} We use network data from the anonymous Facebook100\cite{traud2012social} data set without any associated metadata and for the sole purpose of having realistic network structures from a social media platform. The original data set presents 100 complete and independent networks of Facebook ``friendships'' from 100 American colleges and universities collected as a single-day snapshot in September 2005. Figures \ref{fig:emergence_of_components} and \ref{fig:importance_of_passports} show simulations of our models independently on the 95 Facebook networks with more than 2000 nodes, showing the individual averages obtained from each set of parameters on each network.\\
\textbf{Observability Model.} Our observability model runs on a directed (or undirected\cite{allard2014coexistence}) network of potential data flow where a link from $i$ to $j$ means that user $j$ receives data from user $i$. We simulate an observability process by selecting a fraction $\varphi$ of users who are directly observed by a third-party. For an observability process of depth $L=0$, the simulation is now over and a fraction $\varphi$ of users have been observed. For an observability process of depth $L=1$, all currently unobserved users whose data are received by directly observed users are now also observed (call those users the first generation of indirectly observed users). For an observability process of depth $L=2$, all currently unobserved users whose data are received by the first generation of indirectly observed users are now also observed. While the model can be extended to any depth $l$, Figs. \ref{fig:emergence_of_components} and \ref{fig:importance_of_passports} both use $L=2$. In these figures, we measure the fraction of observed individuals (directly and indirectly observed), the largest component of unobserved individuals connected through uninterrupted data flow, and the fraction of observed individual with a given security setting.\\
\textbf{Distributed Consent Model.} Our distributed consent passport constrains the data flow social media networks and the possibility of users to be directly or indirectly observed. We define a general distributed consent model but implement it using only three distinct security settings: Users at level 0 share their data with all network neighbours and can be directly observed by a third party; users at level 1 share their data with all network neighbours but can \textit{not} be directly observed by a third party; users at level 2 only share their data with neighbours at level 1 or 2 and can \textit{not} be directly observed by a third party. Security levels are randomly assigned to nodes in a network (uniformly at random) at the start of every run of the model with the following probabilities given an adoption frequency $x$ of distributed consent: 2/3 of users are assigned to level 0, 1/3 - $x$ are assigned to level 1, and $x$ are assigned to level 2. In Fig. \ref{fig:emergence_of_components}, a fraction $\varphi = 1\%$ of users at security level 0 are then directly observed and the observability model then runs as defined above but with the directionality of data flow limited by security level 2.\\
\textbf{Consent Passport Model.} We extend the distributed consent passport to a general multilayer network where users are part of multiple platforms. In Fig. \ref{fig:importance_of_passports} we do this by doubling the original network to obtain a two-platform system where social connections exist on two platforms at once. Given an adoption frequency $y$ of consent passport we force a fraction $y$ of the fraction $x$ of adopters of security level 2 to have level 2 on both platforms. On both platform independently, we then distribute uniformly at random the remaining $(1-y)x$ fraction of adopters of level 2 without the passport, then the 1/3-$x$ fraction of level 1 users and the 2/3 fraction of level 0 users. In Fig. \ref{fig:importance_of_passports}, we then select $\varphi = 0.25\%$ of users at security level 0 on at least one platform to be directly observed. The observability model then runs as usual but indirectly observing the network neighbours $j$ of observed node $i$ if data flow from $j$ to $i$ on \textit{at least} one platform.\\
}

\section*{Acknowledgements}


The authors would like to thank James Bagrow for early discussions and feedback on this project. This work is supported by Google Open Source under the Open-Source Complex Ecosystems And Networks (OCEAN) project, MassMutual under the MassMutual Center of Excellence in Complex Systems and Data, the Natural Sciences and Engineering Research Council of Canada, and the Sentinelle Nord program from the Canada First Research Excellence Fund. Any opinions, findings, and conclusions or recommendations expressed in this material are those of the author(s) and do not necessarily reflect the views of the aforementioned financial supporters. 

\section*{Author Contributions Statement}

J.L., R.H. and L.H.-D. conceived the study. J.L. and R.H. developed the conceptual framework. J.O. developed the threat model and cybersecurity framework. A.A. and L.H-D. developed the computational model and produced the results. All authors wrote and reviewed the manuscript.

\section*{Additional Information}

\textbf{Data and code availability} All data and all codes for model implementation and figure replication are available from \url{https://github.com/antoineallard/distributed-consent}.





\end{document}